\documentclass[draftclsnofoot, onecolumn, 12pt]{IEEEtran} 
\IEEEoverridecommandlockouts
\usepackage{cite}
\usepackage{amsmath,amssymb,amsfonts,comment,lipsum}
\usepackage{algorithmic}
\usepackage{graphicx}
\usepackage{textcomp}
\usepackage{xcolor}
\usepackage{hyperref}
\usepackage{algorithm}
\usepackage{setspace} 

\def\BibTeX{{\rm B\kern-.05em{\sc i\kern-.025em b}\kern-.08em
    T\kern-.1667em\lower.7ex\hbox{E}\kern-.125emX}}

\begin{document}

\title{Covering Underwater Shadow Zones using Acoustic Reconfigurable Intelligent Surfaces}

\author{
    \IEEEauthorblockN{
        Longfei Zhao\textsuperscript{1}, Jingbo Tan\textsuperscript{1}, Jintao Wang\textsuperscript{1}, Ian F. Akyildiz\textsuperscript{2}, Zhi Sun\textsuperscript{1§}
    }
    \IEEEauthorblockA{
        \textsuperscript{1}\textit{Department of Electronic Engineering}, Tsinghua University, Beijing, China \\
        Emails: zhaolf23@mails.tsinghua.edu.cn, tanjingbo@tsinghua.edu.cn, wangjintao@tsinghua.edu.cn, zhisun@ieee.org
    }

    \IEEEauthorblockA{
        \textsuperscript{2}\textit{Center for Robotics and Wireless Communications in Challenging Environments, Faculty of Electrical and Computer Engineering}, \\
        University of Iceland IS102 Reykjavik, Iceland \\
        Email: ianaky@hi.is
    }
    \vspace{-11mm}
    \thanks{\textsuperscript{§}Corresponding author: Zhi Sun (zhisun@ieee.org).
    
    }
}

\maketitle
\vspace{-18mm}
\begin{center}
    \fbox{\parbox{\dimexpr\linewidth-2\fboxsep-2\fboxrule\relax}{\centering
    This work has been accepted for publication at IEEE INFOCOM 2025.}}
\end{center}
\begin{abstract}
To better explore and understand the oceans on our planet, the seamless communication coverage of the vast 3D underwater space has been long desired. However, very different from terrestrial networks using radio signals, underwater acoustic communications encounter a unique and severe challenge: nodes in the underwater shadow zones cannot connect to the rest of network even if they are within the LoS (Line of Sight). 
The range of such underwater shadow zones can be tens of kilometers, where no acoustic signal can propagate, causing nodes inside the shadow zones may be disconnected from the network.
Existing solutions  mainly focus on deploying nodes outside of shadow zones. However, such solutions cannot ensure seamless coverage in dynamic ocean environments.
To this end, this paper aims to fundamentally address the shadow zone problem by utilizing the acoustic Reconfigurable Intelligent Surfaces (aRIS) to actively control the underwater channel.
First the shadow zones are modeled analytically, and then based on this model optimal aRIS deployment solutions are developed in order to cover shadow zones in both deep and shallow seas.
Note that the initial aRIS design developed in this paper is  redesigned  by considering practical engineering limitations such that it is implemented on real-world hardware and its performance is validated through pool tests.
The shadow zone coverage performance, evaluated by Bellhop-based simulations, shows that without aRIS deployment, coverage is limited to less than 20\%, regardless of increased radiated energy from the initial coverage source. The ocean environment causes persistent shadow zones, with coverage only in bending regions. However, optimal aRIS deployment achieves nearly 100\% energy coverage across all regions, including previously unreachable shadow zones, by introducing energy into these areas and leveraging the energy-rich regions to enhance shadow zone coverage.




\end{abstract}

\section{Introduction}
The vast and largely unexplored depths of the ocean necessitate
reliable underwater wireless networks to transfer information to and from the underwater, which can support various applications such as environmental monitoring, climate research, archaeological explorations, and biological studies \cite{jiang2023underwater}. 
In contrast to electromagnetic (EM) and optical waves that are severely attenuated in sea water, underwater acoustic communication is the only viable solution to cover the wireless nodes in the vast underwater environment \cite{akyildiz2015realizing}.

Despite the advantages, the complicated underwater environments impose unique challenges in wireless communication and networking \cite{yin2024temporal,10615996}. The underwater shadow zone is one of the unique underwater phenomena that causes severe challenges in the underwater network design. Specifically, the inhomogeneity of the underwater medium, primarily caused by variations in temperature, salinity, and pressure, leads to changes in the sound speed profile (SSP)\cite{stojanovic2009underwater}. 
Those changes cause sound waves to bend, a phenomenon known as refraction. As a result, the shadow zones are formed, where acoustic signals are significantly weakened or completely absent \cite{wang2023dynamic}. 


\begin{figure*}[htbp]
\centerline{\includegraphics[width=1\textwidth]{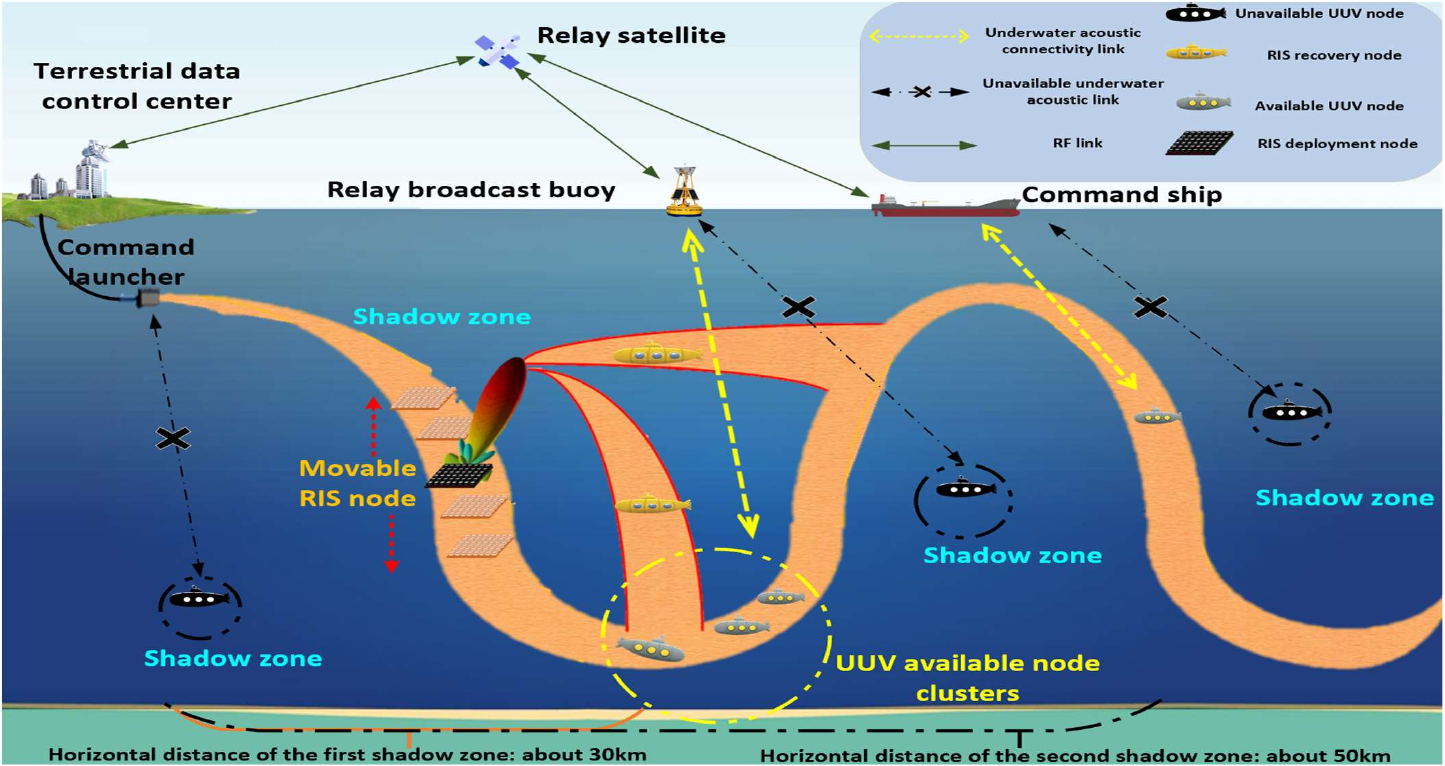}}
\vspace{-1mm} 
\caption{Underwater acoustic communication network coverage.}
\label{fig}
\vspace{-7mm} 
\end{figure*}

Fig. 1 illustrates how the shadow zones impact the underwater network coverage. The envisioned underwater network connects the terrestrial control/data center with all the underwater wireless sensors and Underwater Unmanned Vehicles (UUVs) distributed in a large oceanic region. Due to the unique SSP distribution in the ocean, acoustic waves bend towards the yellow belt in Fig. 1, leaving areas outside the yellow belt with minimal sound energy, i.e. the shadow zone forms \cite{wu2022effects,sandeep2017review}. Any underwater nodes outside the yellow belt are not connected to the network due to the low received signal power. Note that the ratio of the shadow zone is much higher than the yellow belt, which severely affects the coverage performance of underwater acoustic networks \cite{defrianto2023determination, lu2015throughput}.  

To address the aforementioned shadow zone problem, existing solutions focus on positioning underwater nodes to avoid such dark regions, allowing nodes to bypass shadow zones for better coverage \cite{domingo2009optimal, huang2020adaptive}. However, as shown in Fig. 1, the area of shadow zones is huge while the yellow belt is much narrower, leaving very small room to bypass shadow zones.
More importantly, for applications that the target area is inside the shadow zone, the wireless nodes have to be deployed there. Then it is impossible to complete the mission by just passively adapting the shadow zone distribution. Hence, instead of the existing passive shadow zone avoidance solutions, a more reliable solution which can actively cover the shadow zone is essential to realize the seamless underwater communication coverage. Unfortunately, to the best of our knowledge, no existing solution has been reported towards the active shadow zone coverage problem in the literature.


In this paper, first time in the literature we propose to use acoustic Reconfigurable Intelligent Surface (aRIS) \cite{sun2022high} to proactively cover the underwater shadow zone. Though terrestrial RIS has been intensively investigated for 6G mobile networks \cite{zhang2022active}, aRIS has only been proposed recently \cite{sun2022high, wang2023designing}. Here we introduce  a brand new way to utilize the aRIS that addresses the aforementioned unique underwater networking challenge, namely to cover the shadow zones. In particular, as shown in Fig. 1, aRIS can be strategically positioned (e.g., inside the yellow belt in Fig. 1) and can reflect acoustic signals towards shadow zones, thereby increasing the overall underwater network coverage. 
The proposed aRIS-aided shadow zone coverage solution is applicable for both deep and shallow sea environments. We demonstrate how aRIS can achieve unprecedented underwater coverage efficiency.
The contributions of our paper can be summarized as follows.
\begin{itemize}
\item \textbf{Analytical modelling of shadow zones in marine environments:}
We first derive a comprehensive model of shadow zones in marine environments based on the ray-acoustics theory, using the sound channel axis depth to distinguish deep-sea and shallow-sea conditions. The established shadow zone model can help us to quantitatively evaluate the area of shadow zones, which lays the foundation of determining  the optimal positions to deploy the aRIS.



\item \textbf{Improved design of acoustic RIS considering practical constraints:}
The initial design of aRIS is difficult to implement due to practical considerations, such as the inability of piezoelectric elements to simultaneously perform forward and reverse effects\cite{sherman2007transducers}, necessitating separate units for transmission and reception\cite{shen2019digital}. Our novel aRIS design addresses these limitations by using two piezoelectric elements working in tandem\cite{butler2018properties}, operating on a "first absorb, then radiate" mechanism to obtain high beamforming gain\cite{zhang2022active}. This configuration enables precise control over sound ray angles, facilitating targeted coverage. A prototype has validated the practical feasibility and effectiveness of our approach in a pool environment.


\item \textbf{Optimal aRIS deployment to cover shadow zones:}
To address the unique challenges in both deep and shallow-sea environments, we propose an aRIS-aided shadow zone coverage solution. The solution theoretically introduces optimal deployment of aRIS to enhance the shadow zone coverage. In the deep-sea environment, deploying the aRIS at the sound channel axis is proven to be optimal by exploiting the natural acoustic properties of the marine environment. For the shallow-sea environments, we provide a quantitative relationship between the shadow zone area and the aRIS position, and accordingly propose an optimization method for aRIS placement that maximizes the coverage area. The proposed solution is validated through rigorous simulations and theoretical analysis, by demonstrating its ability to adapt to diverse oceanic conditions, offering a robust solution for both deep-sea and shallow-sea coverage applications.

\item \textbf{Impact analysis of dynamic marine environment and corresponding robust solution:}
We analyze the impact of dynamic marine environments on aRIS coverage, particularly focusing on how variations in aRIS platform behavior may influence the coverage areas. Our comprehensive theoretical models and derivations account for measurement errors and dynamic conditions, ensuring a robust aRIS deployment and effective reflection performance. Our intelligent underwater aRIS, capable of sensing displacement and rotation, adapts to environmental changes by performing real-time compensation of the reflection phase. It ensures stable aRIS coverage, maintaining effective shadow zone coverage and enhancing network resilience. This adaptability guarantees consistent and reliable underwater acoustic communication, even in fluctuating marine conditions, providing a robust solution for real-world scenarios.

\end{itemize}


\section{Shadow Zone Models in Deep-Sea and Shallow-Sea Environments}\label{Model}
\begin{figure}[htbp]
\centerline{\includegraphics[width=1\textwidth]{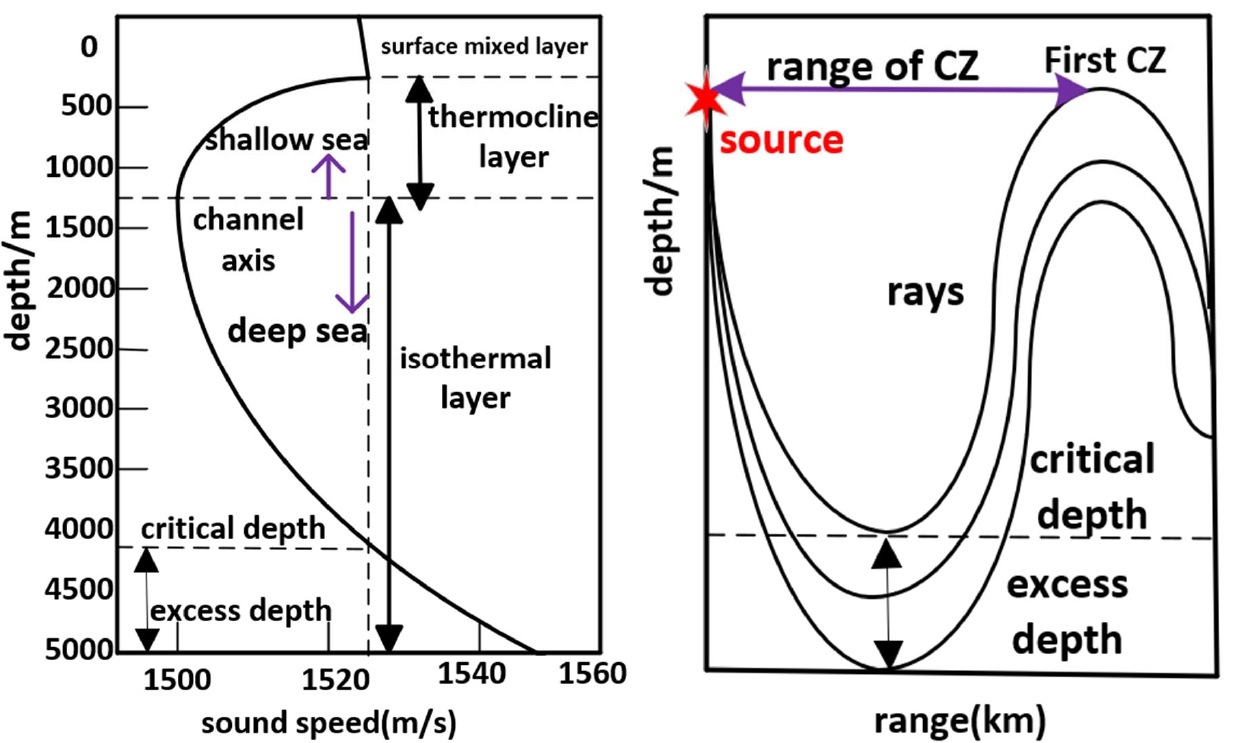}}
\caption{Standard ocean sound speed profile and layered structure (left); Deep-sea V-shaped coverage pattern (right). Sound rays tend to bend toward regions with lower sound speed. As a result, the sound rays above the sound channel axis bend downward, while those below the axis bend upward, forming the V-shaped propagation path shown in the figure.}
\label{fig:SSPandv}
\vspace{-1mm}
\end{figure}

In practical oceanic environments, the sound speed $c(x, y, z)$ is inherently three-dimensional, varying with spatial position. However, in many cases, the horizontal variations in sound speed (across $x$ and $y$) are relatively small when compared to the dominant variation with depth ($z$). This allows us to reasonably approximate the sound speed as a one-dimensional function of depth, $c(z)$. Therefore, we model shadow zones in deep-sea and shallow-sea environments based on SSPs\cite{munk1983ocean}.

\begin{equation}
c(z) = c_0 \left[ 1 + \epsilon \left( \frac{z - z_0}{z_s} + e^{-\frac{z - z_0}{z_s}} - 1 \right) \right],
\end{equation}


\noindent where \(c(z)\) is the sound speed at depth \(z\), \(c_0\) is the sound speed at the reference depth \(z_0\), \(\epsilon\) is a small parameter, which is typically around 0.00737 and describes the amplitude of sound speed variation, \(z_0\) is the depth of the sound channel axis, which is typically 2100 meters, and \(z_s\) is the scale depth, which is typically 1300 meters.
As shown on the left side of Fig. \ref{fig:SSPandv}, the standard ocean SSP can be categorized into deep-sea and shallow-sea scenarios based on sound speed minima or a sound channel axis\cite{hui2022underwater}. In deep-sea scenarios, the sound speed decreases to a minimum at the sound channel axis depth before increasing with further depth, creating a V-shaped coverage pattern with convergence zones (CZs) and shadow zones. Conversely, in shallow-sea scenarios, the surface layer causes sound speed to first increase and then decrease with depth, leading to different shadow zone manifestations.

\subsection{Deep-Sea Shadow Zone Model}
The SSP in the deep-sea environment creates a V-shaped coverage pattern with CZs and shadow zones\cite{lawrence1983simple}, as shown on the right side of Fig. \ref{fig:SSPandv}. For a given initial grazing angle \(\theta_0\), the horizontal propagation span \(r\) is:
\begin{equation}
r = \int_{z_0}^{z} \frac{\cos \theta_0}{\sqrt{n^2(z) - \cos^2 \theta_0}} dz, \label{eq}
\end{equation}
where \(n(z)=\frac{c_0}{c(z)}\). Therefore, the coverage area \(A_{\text{coverage}}\) is:
\begin{equation}
A_{\text{coverage}} = \left\{ (r, z) \ \middle| \ r_{\min}(z) < r < r_{\max}(z) \right\}, \label{eq}
\end{equation}
\begin{subequations}
\begin{equation}
r_{\max} = \int_{z_0}^{z} \frac{\cos \theta_{\min}}{\sqrt{n^2(z) - \cos^2 \theta_{\min}}} \, dz, \label{eq:a}
\end{equation}
\begin{equation}
r_{\min} = \int_{z_0}^{z} \frac{\cos \theta_{\max}}{\sqrt{n^2(z) - \cos^2 \theta_{\max}}} \, dz. \label{eq:b}
\end{equation}
\end{subequations}
where $\theta_{\max}$ and $\theta_{\min}$ define the angular bounds for the grazing angle of the emitted acoustic source relative to the horizontal plane. The shadow zones \(A_{\text{shadow}}\) are:
\begin{equation}
A_{\text{shadow}} = A_{\text{coverage}}^c = \left\{ (r, z) \ \middle| \ r \notin \left( r_{\min}(z), r_{\max}(z) \right) \right\}.
\end{equation}
The coverage and shadow zone areas are quantified as:
\begin{equation}
S_{\text{coverage}} = \iint_{A_{\text{coverage}}} \, dr \, dz, \label{eq:coverage_area}
\end{equation}
\begin{equation}
S_{\text{shadow}} = \iint_{A_{\text{shadow}}} \, dr \, dz = S_{\text{deep-sea}} - S_{\text{coverage}}, \label{eq:shadow_area1}
\end{equation}
where \(S_{\text{deep-sea}}\) represents the whole area that we are focusing on. In this paper, the \(S_{\text{deep-sea}}\) is defined as the area including the first bending region of sound rays.
Given the well-behaved nature of \(n(z)\) together with the bounded and continuous functions \(\theta_{\min}(z)\) and \(\theta_{\max}(z)\), these integrals in (\ref{eq:coverage_area}) and (\ref{eq:shadow_area1}) are continuous and bounded. Thus, we can quantitatively determine the coverage and shadow zone areas, providing a precise measure of underwater acoustic communication coverage effectiveness.

\begin{figure}[htbp]
\centerline{\includegraphics[width=1\textwidth]{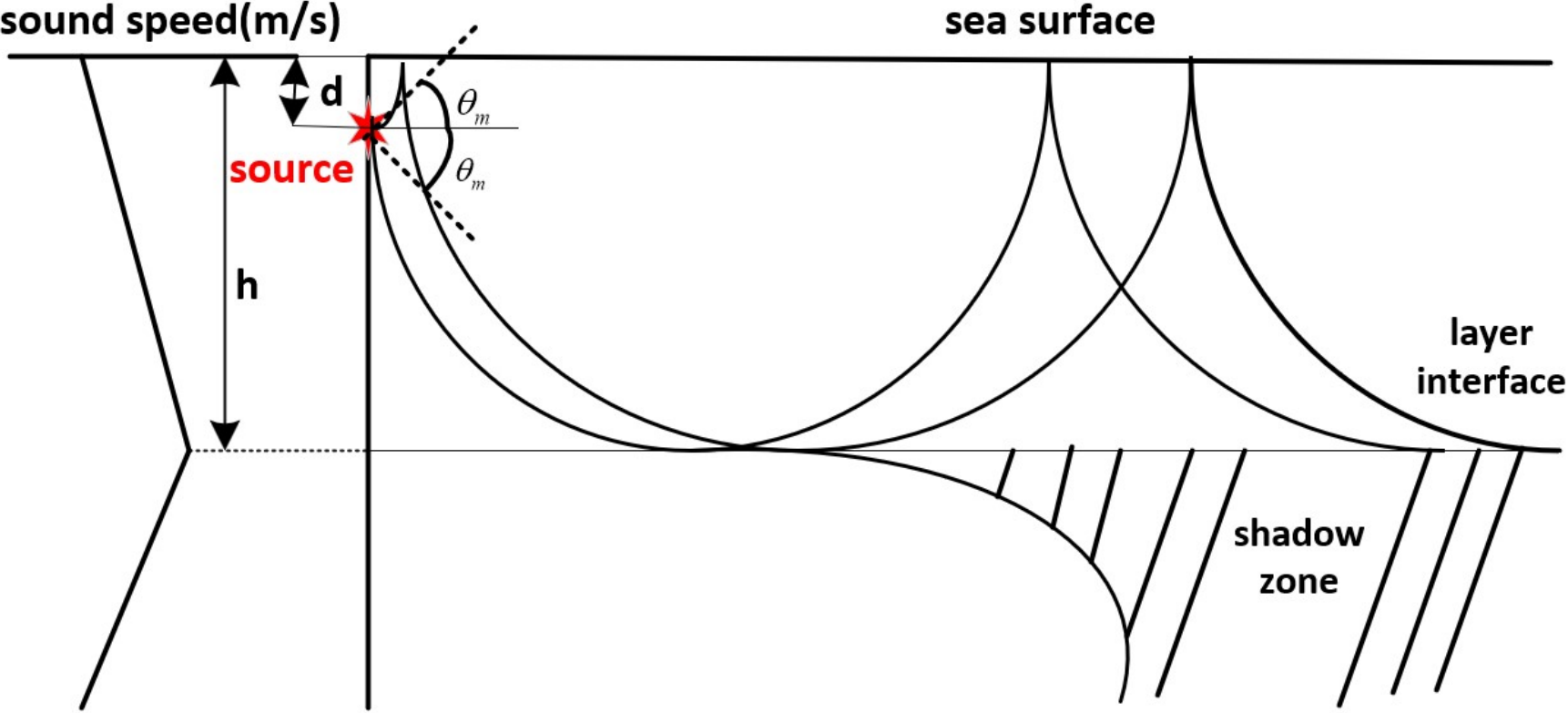}}
\caption{Shallow-sea surface waveguide.}
\label{sh}
\vspace{-1mm}
\end{figure}
\vspace{-5mm}

\subsection{Shallow-Sea Shadow Zone Model}
In shallow-sea environments, the surface mixed layer with a positive sound speed gradient forms an upper channel boundary. The critical grazing angle \(\theta_m\) satisfies Snell's law:
\begin{equation}
\frac{\cos \theta_m}{c_0} = \frac{1}{c_h},
\end{equation}
where \(c_0\) is the sound speed at the source depth, and \(c_h\) is the sound speed at the surface mixed layer depth. Sound rays emitted at angles larger than \(\theta_m\) form shadow zones below this depth, creating a surface waveguide, as shown in Fig. \ref{sh}. The primary coverage area is confined within the surface waveguide depth. Below this layer, sound rays bend towards the seabed and are absorbed, preventing energy coverage in the lower layers. Therefore, the shadow zone \(A_{\text{shadow}}\) is defined as:
\vspace{-2mm}
\begin{equation}
A_{\text{shadow}} = \left\{ (r,z) \ \middle| \ h<z<D \text{ and } r > r_{\min}(z) \right\}, \label{eq:shadow}
\end{equation}
where \(h\) is the waveguide depth, and \(D\) is the total depth. The coverage area \(A_{\text{coverage}}\) is:
\begin{equation}
A_{\text{coverage}} = \left\{ (r,z) \ \middle| \ 0<z<h \text{ and } 0 \le r \le r_{\max}(z) \right\}, \label{eq:coverage}
\end{equation}

The shadow zone area \(S_{\text{shadow}}\) is:
\begin{equation}
S_{\text{shadow}} = \int_{h}^{D} \int_{r_{\min}(z)}^{R_{\text{max}}} dr \, dz. \label{eq:shadow_area2}
\end{equation}

By analyzing the physical propagation characteristics in both deep-sea and shallow-sea environments, we have established a clear model of the coverage and shadow zones inherent to physical parameters. This shadow zone model can help us to explore the strategy to mitigate shadow zones by utilizing aRIS, and thus enhance the overall effectiveness of underwater acoustic communication systems.
The detailed methodologies and strategies for the optimal deployment of aRIS will be discussed in Section \ref{RISDeploy}.

\section{Improving the Current Underwater Acoustic RIS Design Considering Practical Constraints}\label{RISDesign}

To realize the coverage enhancement by utilizing aRIS, a practical design of acoustic RIS is necessary. Therefore, in this section, we first introduce an improved and practical aRIS design solving practical deployment challenges, and then we deploy them in pool tests and evaluate their performance.

\subsection{Underwater Acoustic RIS Hardware Design}\label{AA}
The theoretical model of the aRIS assumes ideal conditions where each reflective unit can independently manipulate the phase and amplitude of incoming acoustic waves\cite{sun2021acoustic}. However, practical implementation faces significant challenges, primarily due to the nature of the piezoelectric effect. The forward and reverse piezoelectric effects cannot perform simultaneously\cite{sherman2007transducers}, necessitating separate units for transmitting and receiving signals in full-duplex acoustic communication systems\cite{shen2019digital}.
To bridge the gap between the ideal model and practical implementation, we introduce a novel design approach. Each reflective unit in our design comprises two piezoelectric elements that work in tandem, as shown in Fig. 4. These elements operate on a "first absorb, then radiate" mechanism. The structure of a single piezoelectric element is shown on the bottom-left of Fig. 4.

\begin{figure}[htbp]
\centerline{\includegraphics[width=1\textwidth]{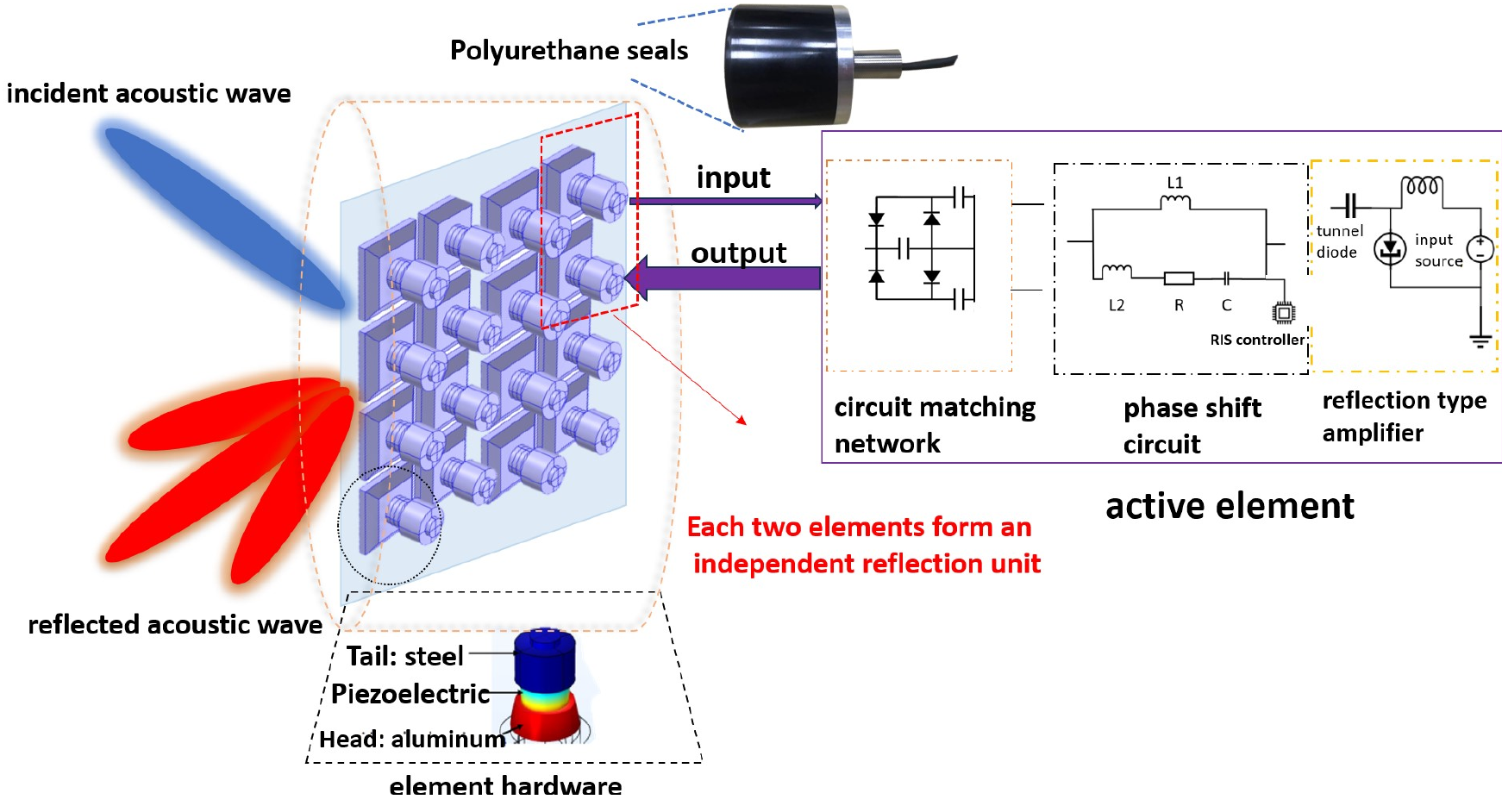}}
\caption{Illustration of the underwater acoustic RIS hardware.}
\label{fig}
\end{figure}
The piezoelectric element used here has the same structure as the Tonpilz transducer, commonly employed in existing underwater acoustic communication systems\cite{sun2022high}. Our design utilizes one piezoelectric element as an absorber and the other as a radiator. The received acoustic waves are first absorbed and converted into electrical signals, which are then processed through amplifiers and phase shifters to control the amplitude and phase of the re-radiated waves. Together, the two elements form a "reflection unit." Multiple reflection units are assembled into an array to form the designed aRIS. As illustrated in Fig. 5, by controlling the phase of each reflection unit after absorbing the incoming signal, the aRIS can re-radiate the signal in a controlled manner, enabling high-gain beamforming\cite{cai2020practical}, which can enhance the shadow zone coverage\footnote{For detailed methodologies on underwater acoustic beamforming, refer to the comprehensive analysis in \cite{wang2023designing}}. Unlike amplify-and-forward (AF) relays, the piezoelectric transducer-based aRIS reflects incident waves immediately through the intrinsic piezoelectric effect without needing signal receiving, processing, and re-transmitting modules.

\subsection{Hardware Prototypes and Pool Tests}\label{BB}
To validate our proposed design, we conduct a series of experiments in an underwater environment. The experiments are performed in a pool to test the reflection and beamforming capabilities of the aRIS. The reflection angles are set to 0 degrees for simplicity, with incident angles as 30 and 60 degrees.
\vspace{-3mm}
\begin{figure}[htbp]
\centerline{\includegraphics[width=1\textwidth]{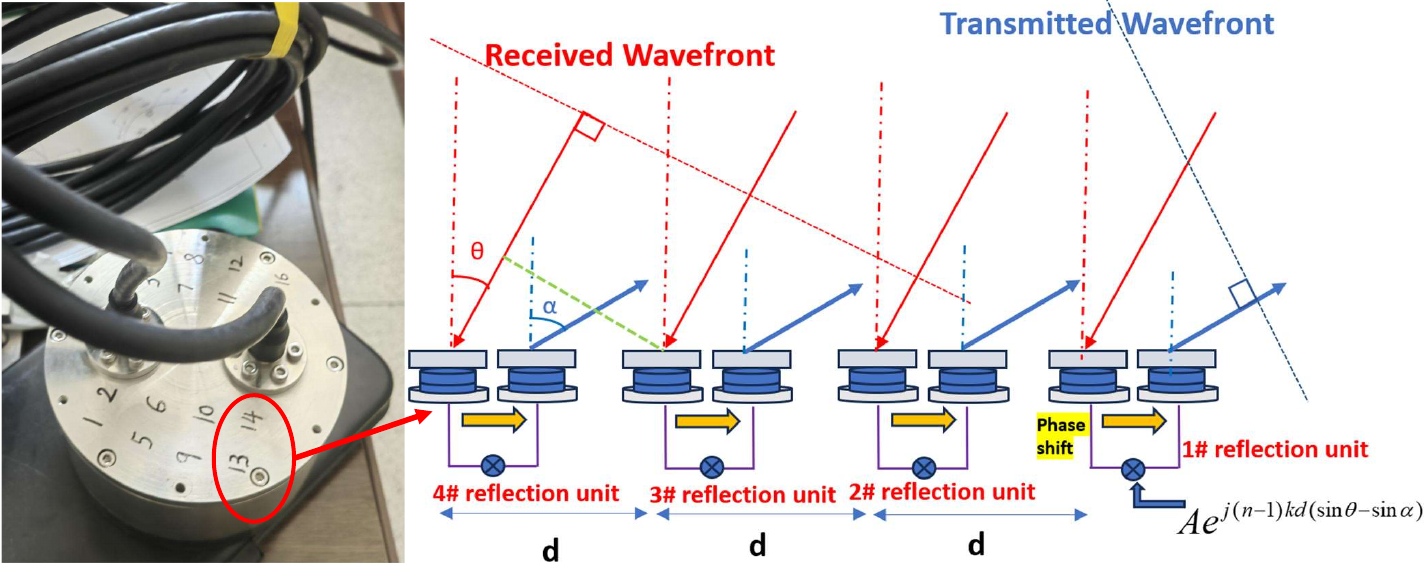}}
\vspace{-3mm} 
\caption{Wet-end setup of the RIS experiment.}
\label{fig:wet_end}
\vspace{-4mm} 
\end{figure}
\vspace{-4mm}
\begin{figure}[htbp]
\centerline{\includegraphics[width=1\textwidth]{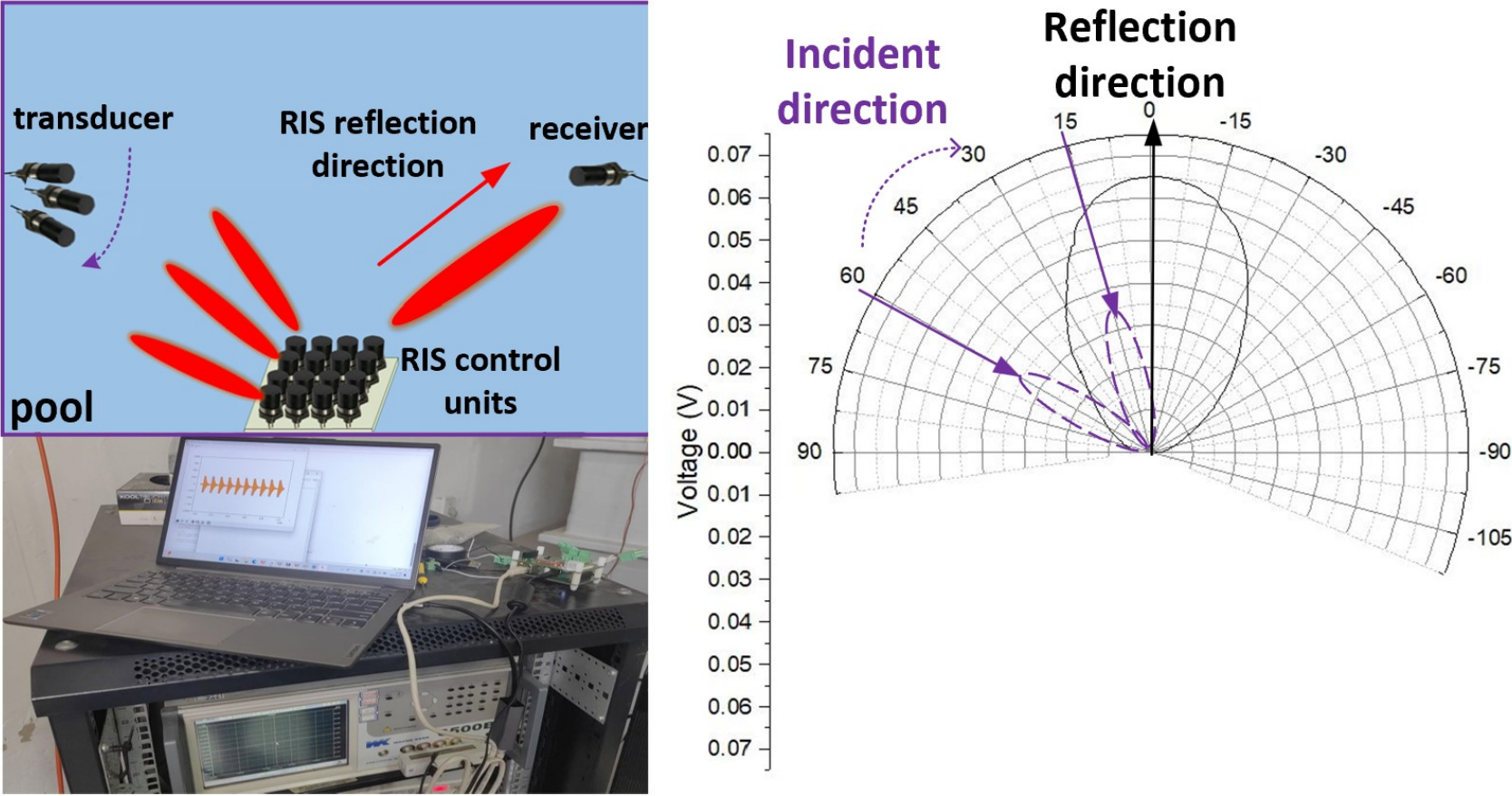}}
\vspace{-3mm} 
\caption{Pool Tests experiment setup (left); Measured beam pattern of the reflected acoustic waves (right).}
\label{fig:water_tank}
\vspace{-2mm} 
\end{figure}

The beam pattern measured during the experiments is shown on the right side of Fig. \ref{fig:water_tank}. 
These results verify the correctness of our hardware design, indicating that the aRIS can effectively accept incoming acoustic waves at various angles and reflect them in a controlled manner.
For this hardware validation, we used two reflection units. The theoretical beamforming gain is $10\log(N)=10\log(2)$ which is approximately $3$dB ( $N$ is RIS units  number). 
The reflected beamforming gain in the measured beam pattern is approximately 2.69 dB, which is consistent with the theoretical results. This calculation is based on comparing the voltage amplitude at the 0-degree reflection direction with the voltage amplitude at the incident angle.
The practical hardware design and its successful experimental validate the feasibility of implementing the designed aRIS to enhance shadow zone coverage.


\section{Optimal Deployment of Underwater Acoustic RIS for Shadow Zone Coverage}\label{RISDeploy}
In this section, we propose the aRIS aided shadow zone coverage solution based on the established shadow zone model in both deep-sea and shallow-sea environments. Sec. IV-A focuses on analyzing the effectiveness of deploying RIS for deep-sea environments. Sec. IV-B delves into the optimization problem of RIS deployment to enhance shallow-sea coverage.

\subsection{Deep-Sea Acoustic RIS Array Coverage}\label{AA}
In the deep-sea environment, the sound propagation is influenced by the SSP and boundaries such as the sea surface and seabed\cite{gul2017underwater}. When an aRIS is deployed, the reflection angles can lead to various propagation paths. Our objective is to mitigate the shadow zones and enhance the coverage by focusing on rays that bend within the water column, avoiding energy loss at the boundaries. These rays are analogous to the sound fixing and ranging (SOFAR) channel propagation in the deep sea, where the sound speed minimum traps sound waves, allowing long-distance travel with minimal attenuation.
Given this understanding, the optimal deep-sea RIS deployment is proved by \textit{Theorem 1}.

\newtheorem{thm}{Theorem}
\begin{thm}
For deep-sea coverage, the optimal depth \( h_{\text{opt}} \) to deploy the aRIS is at the sound channel axis depth \( H \). The coverage gain \(\xi\) when deploying the aRIS at the sound channel axis depth \( H \) compared with deploying the aRIS at arbitrary shallower depth $h$ is given by
\begin{equation}
\xi=\frac{r_{\text{max}, H}}{r_{\text{max}, h}}-1.
\end{equation}
\end{thm}

\noindent\textit{Proof}. Here, we equate an aRIS to a source, and thus the coverage realized by the aRIS can be analyzed based on the established coverage model. Specifically, define the coverage efficiency \(\eta\) as the ratio of the coverage area to the shadow zone area. According to (\ref{eq:coverage_area}) and (\ref{eq:shadow_area1}), \(\eta\) satisfies
\begin{equation}
\eta = \frac{\text{S}_{\text{coverage}}}{\text{S}_{\text{shadow}}}=
\frac{\int_{0}^{D} \int_{r_{min}}^{r_{max}}dz \, dr}{\text{S}_{\text{deep-sea}}-\int_{0}^{D} \int_{r_{min}}^{r_{max}}dz \, dr}.
\label{xi}
\end{equation}
Since the integrals over \( z \) have the same limits, the comparison focuses on the horizontal span \( r \). Increasing \( r\) effectively reduces the shadow zone area. Therefore, the optimal deployment of the aRIS is to maximize the horizontal span \( r \) of the coverage area. Due to the symmetry of the deep-sea sound channel, placing the aRIS at depth \( h>H \) is less efficient compared to placing it at the symmetric depth \( 2H - h \). Therefore, we only consider to deploy the aRIS at or above \( H \).
Using Snell's law at the interface between the sea surface and the sound source, we have
\begin{equation}
\frac{\cos \theta_h}{c_h} = \frac{\cos \theta_H}{c_H} = \frac{1}{c_s},
\label{eq}
\end{equation}
where \(\theta_h\) is the grazing angle at depth \( h \) with $h<H$, \(\theta_H\) is the grazing angle at depth \( H \), \( c_h \) is the sound speed at depth \( h \), \( c_H \) is the sound speed at depth \( H \), and \( c_s \) is the sound speed at the sea surface.
The sound rays emitted from the aRIS that tangentially intersect the sea surface form the boundary of the V-shaped coverage area. Utilizing the beamforming capabilities of the aRIS, the entire area within the V-shaped boundary is effectively covered. Therefore, a larger V-shaped span implies a greater \( r_{\text{max}} - r_{\text{min}} \) at each depth, corresponding to a larger coverage area. Consequently, we only need to consider the span of the V-shaped boundary, which is defined by the sound rays emitted from the RIS that tangentially intersect the sea surface, corresponding to \( r_{\text{max}} \).

Given the periodicity of sound ray spans, the effective span is determined by the shortest period. The maximum spans for sound rays reflected from depths \( H \) and \( h \) are
\begin{subequations}
\begin{equation}
r_{\text{max}, H} = \int_{0}^{H} \frac{\cos \theta_H}{\sqrt{n_H^2 - \cos^2 \theta_H}} \, dz
= \int_{0}^{H} \frac{\frac{c_H}{c_s}}{\sqrt{\frac{c_H^2}{c^2(z)}- \frac{c_H^2}{c_s^2}}} \, dz,
\label{eq:a}
\end{equation}
\begin{equation}
r_{\text{max}, h} = \int_{0}^{h} \frac{\cos \theta_h}{\sqrt{n_h^2 - \cos^2 \theta_h}} \, dz
=\int_{0}^{h} \frac{\frac{c_h}{c_s}}{\sqrt{\frac{c_h^2}{c^2(z)} - \frac{c_h^2}{c_s^2}}} \, dz.
\label{eq:b}
\end{equation}
\end{subequations}
Thus, the ratio of these spans can be presented as
\begin{equation}
\frac{r_{\text{max}, H}}{r_{\text{max}, h}} = \frac{c_H}{c_h} \left( \frac{\int_{0}^{h} \frac{1}{\sqrt{\frac{c_H^2}{c^2(z)} - \frac{c_H^2}{c_s^2}}} \, dz + \int_{h}^{H} \frac{1}{\sqrt{\frac{c_H^2}{c^2(z)} -\frac{c_H^2}{c_s^2}}} \, dz}{\int_{0}^{h} \frac{1}{\sqrt{\frac{c_h^2}{c^2(z)} - \frac{c_h^2}{c_s^2}}} \, dz} \right).
\label{span}
\end{equation}
Simplifying (\ref{span}) and letting \(\xi = \frac{\int_{h}^{H} \frac{1}{\sqrt{\frac{1}{c^2(z)} - \frac{1}{c_s^2}}} \, dz}{\int_{0}^{h} \frac{1}{\sqrt{\frac{1}{c^2(z)} - \frac{1}{c_s^2} }} \, dz}\), we have
\begin{equation}
\frac{r_{\text{max}, H}}{r_{\text{max}, h}} = 1 + \frac{\int_{h}^{H} \frac{1}{\sqrt{\frac{1}{c^2(z)} - \frac{1}{c_s^2}}} \, dz}{\int_{0}^{h} \frac{1}{\sqrt{\frac{1}{c^2(z)} - \frac{1}{c_s^2} }} \, dz} = 1 + \xi,
\end{equation}
which proves (\ref{xi}). Since \(c(z) < c_s\) for all \(z\) in the range \(0 \le z \le H\), both the numerator and denominator of \(\xi\) are positive, ensuring \(\xi > 0\). Thus, we have
\begin{equation}
\frac{r_{\text{max}, H}}{r_{\text{max}, h}} > 1,
\end{equation}
which indicates to deploy the acoustic RIS at $H$ obtains maximum span of V-shape propagation pattern, leading to the optimal coverage area.\hfill$\blacksquare$

\textit{Theorem 1} demonstrates that deploying the aRIS at \(H\) is optimal to enhance shadow zone coverage. Hence, we propose to deploy the aRIS at the sound channel axis depth in the deep-sea environment. Moreover, for the complementary RIS placed at the sound channel axis depth on the opposite side of the V-shaped propagation pattern, the same principle holds due to the symmetric nature of sound ray propagation.

\subsection{Shallow-Sea aRIS Wide-Area Coverage}\label{BB}
In the shallow-sea environment, the sound speed variation is relatively small compared to the deep-sea environment. Therefore, it is reasonable to approximate the SSP with a linear gradient, modeling the sound rays as circular arcs due to the nearly constant gradient in sound speed\cite{bass2003physical}. For signals reflected by the aRIS, these sound rays are modeled as circular arcs to analyze their coverage capability\cite{wang2018linear}. To reduce the shadow zone area, as shown in (\ref{eq:shadow_area2}), it is essential to increase the minimum horizontal range \( r_{\min}(z) \), corresponding to the sound ray reaching the sea-bed after leaving the surface layer. Inspired by this, we propose deploying the aRIS below the surface layer, where sound rays with large grazing angles penetrate, to mitigate shadow zones in the lower layers. As shown in Fig. \ref{fig:shallow_sea_coverage}, the optimal aRIS deployment depth should maximize the coverage distance below the surface waveguide. To this end, we formulate an optimization problem to determine this optimal depth and provide the corresponding solution.

\begin{figure}[htbp]
\centerline{\includegraphics[width=1\textwidth]{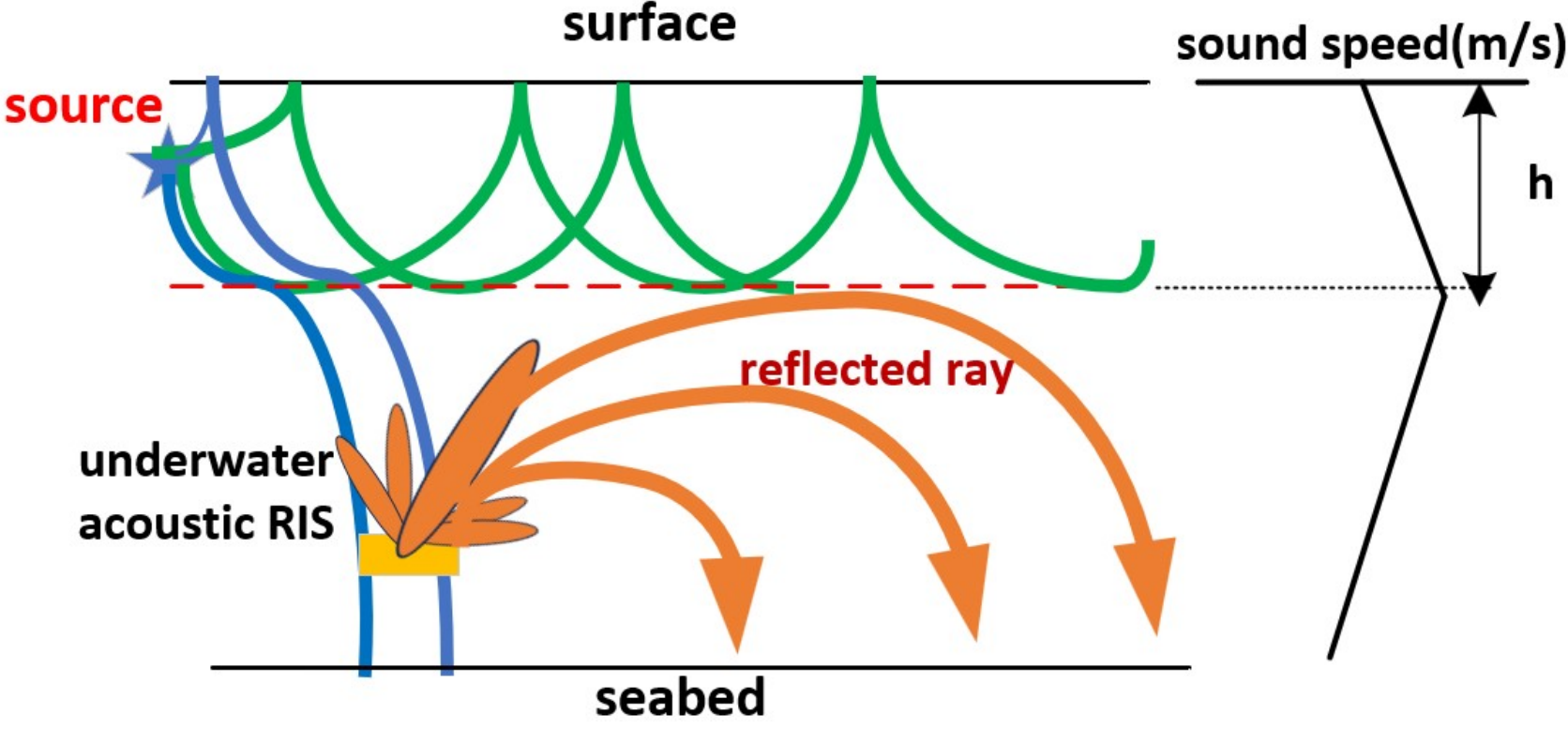}}
\caption{Shallow Sea RIS coverage.}
\label{fig:shallow_sea_coverage}
\end{figure}

Specifically, the coverage distance $r$ can be expressed based on the circular arcs approximation of the sound ray as
\vspace{-2mm}
\begin{equation}
r \triangleq \frac{|z_{\text{RIS}} - h|}{\tan \left( \frac{\theta_{\text{reflect}}}{2} \right)} + \frac{|D - h|}{\tan \left( \frac{\theta_{D}}{2} \right)},
\label{eq:shallow_r}
\end{equation}
where \( z_{\text{RIS}} \) is the aRIS depth, \( h \) is the surface waveguide depth, \( D \) is the total depth, \( \theta_{\text{reflect}} \) is the reflection grazing angle at the aRIS depth, and \( \theta_{D} \) is the grazing angle at the seabed. Then, the aRIS deployment optimization problem can be formulated as
\begin{equation}
\begin{aligned}
    & \text{maximize } r  \\
    & \text{subject to } \theta_{\text{reflect}} = \arccos \left( \frac{c(z_{\text{RIS}})}{c_h} \right), \\ 
    & \ \qquad\qquad\theta_D = \arccos \left( \frac{c_D \cos (\theta_{\text{reflect}})}{c(z_{\text{RIS}})} \right).
\end{aligned}
\label{eq:shallow_optimization}
\end{equation}
where \( c(z_{\text{RIS}}) \) is the sound speed at the aRIS depth, \( c_h \) is the sound speed at the surface waveguide depth, and \( c_D \) is the sound speed at the seabed.

Given the non-convex nature of the problem in (\ref{eq:shallow_optimization}), we apply numerical optimization techniques such as Sequential Quadratic Programming (SQP) \cite{boggs1995sequential} or heuristic methods like Genetic Algorithms (GA)\cite{jaramillo2002use} to find the optimal solution.

\begin{algorithm}[H]
\caption{Numerical Optimization for aRIS Placement}
\begin{algorithmic}[1]
\STATE \textbf{Input:} Depth range \( z_{\text{RIS}} \), sound speed profile \( c(z) \), sound speed \( c_h \) and \( c_D \).
\STATE \textbf{Initialization:} Initialize \( z_{\text{RIS}} \).
\WHILE{not converged}
    \STATE Calculate \( \theta_{\text{reflect}} = \arccos \left( \frac{c(z_{\text{RIS}})}{c_h} \right) \)
    \STATE Calculate \( \theta_D = \arccos \left( \frac{c_D \cos (\theta_{\text{reflect}})}{c(z_{\text{RIS}})} \right) \)
    \STATE Evaluate \( r = \frac{|z_{\text{RIS}} - h|}{\tan \left( \frac{\theta_{\text{reflect}}}{2} \right)} + \frac{|D - h|}{\tan \left( \frac{\theta_{D}}{2} \right)} \)
    \STATE Update \( z_{\text{RIS}} \) using the chosen optimization method
\ENDWHILE
\STATE \textbf{Output:} Optimal \( z_{\text{RIS}}^{*} \)
\end{algorithmic}
\end{algorithm}
By applying these numerical optimization techniques, we can obtain the optimal aRIS placement depth \( z_{\text{RIS}}^{*} \). The specific optimal solutions and detailed results of the numerical analysis are presented in Section V.

\section{Validation And Evaluations}\label{Sim}
In this section, we utilize Bellhop-based simulations to validate the performance of the proposed aRIS aided shadow zone coverage solution in an underwater acoustic field\cite{Porter2021}. We compare the scenarios with and without aRIS assistance to demonstrate the effectiveness of aRIS in enhancing the coverage and filling shadow zones.

\subsection{Deep-Sea Coverage Enhancement with Acoustic RIS}\label{AA}

\begin{figure}[htbp]
\centerline{\includegraphics[width=1\textwidth]{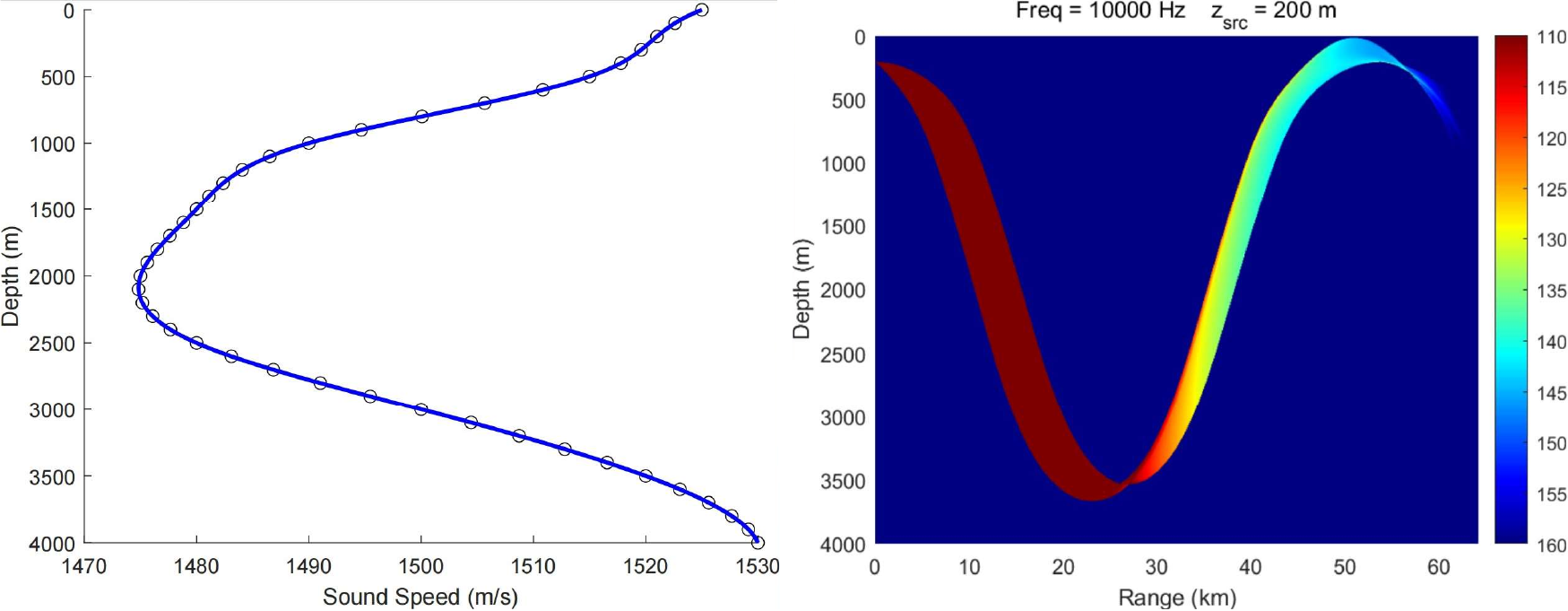}}
\caption{Standard Munk deep-sea sound speed profile model (left); Deep-sea coverage model showing V-shaped structure (right).}
\label{fig:deepsea_coverage}
\vspace{-2mm} 
\end{figure}

To facilitate practical verification, we consider a single-carrier underwater acoustic communication network operating at 10 kHz, using the standard Munk deep-sea sound speed profile model\cite{munk1983ocean}. The sound speed distribution is shown on the left side of Fig. \ref{fig:deepsea_coverage}. The deep-sea environment has a depth of 4 km, with the sound channel axis at 2100 m and the sound source depth at 200 m. Using grazing angles from 0° to 4°, we initiate broadcast communication and focus on the first shadow zone within a single hop. We utilize the Bellhop numerical simulation model to calculate the energy distribution at spatial points, plotting the relative energy magnitude across the space. \footnote{The values shown in the colormap represent transmission loss in decibels (dB), defined as \(TL = -10 \log_{10} (P / P_0)\), where \(P\) is the received energy and \(P_0\) is the reference energy at the source. Higher TL values indicate greater energy attenuation. The colorbar is labeled accordingly as "Transmission Loss (dB)".} As illustrated on the right side of Fig. \ref{fig:deepsea_coverage}, the deep-sea coverage model exhibits a V-shaped structure with distinct shadow and convergence zones. The observed V-shaped energy distribution matches well with our theoretical model, confirming the presence of both coverage and shadow zones as defined in our sets \(A_{coverage}\) and \(A_{shadow}\).

\begin{figure}[htbp]
\centerline{\includegraphics[width=1\textwidth]{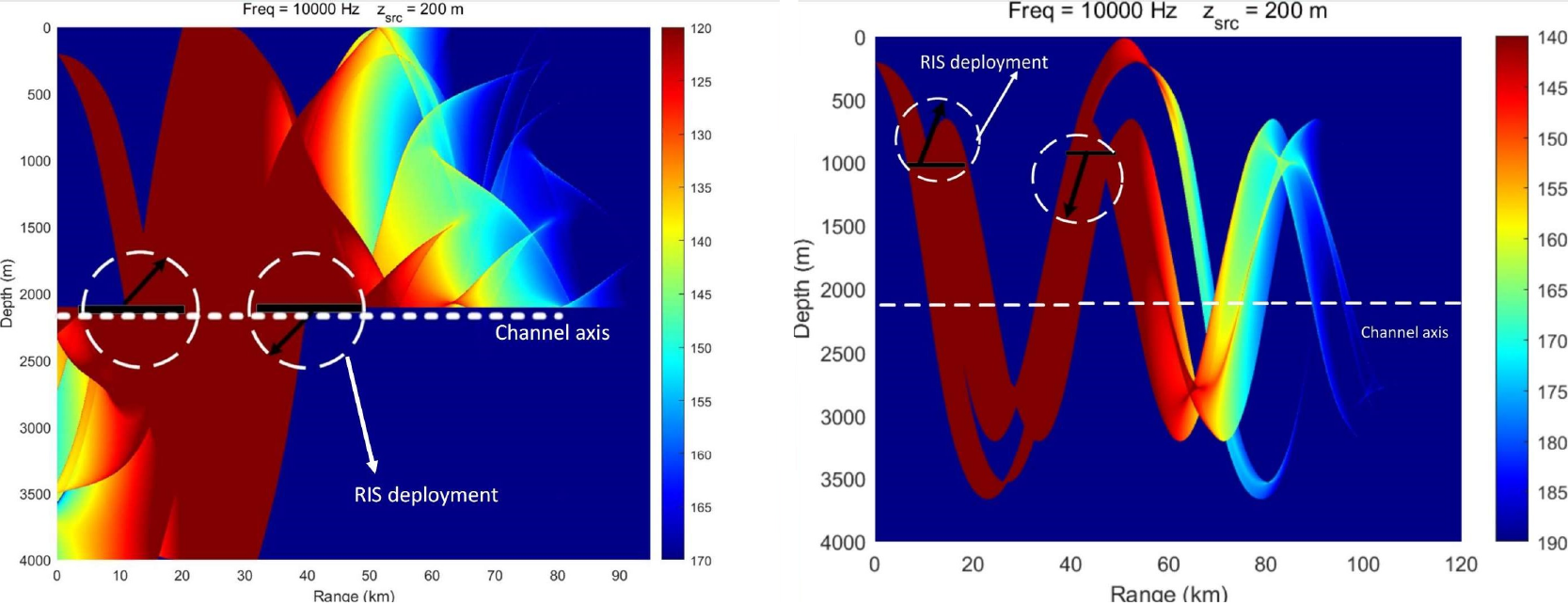}}
\vspace{-3mm} 
\caption{Deep-Sea Acoustic Dual RIS Array Optimal Coverage(left);Deep-Sea Acoustic Dual RIS Array Non-optimal Coverage(right).}
\label{fig:deepsea_op_coverage}
\vspace{-3mm} 
\end{figure}

We test aRIS deployment at various depths to analyze their impact on shadow zone coverage. As shown in Figure \ref{fig:deepsea_op_coverage} (right), non-optimal aRIS placements away from the sound channel axis leave residual shadow zones. Deploying two aRIS units symmetrically at the sound channel axis depth (2100 m) (left side of Fig. \ref{fig:deepsea_op_coverage}) effectively redirects energy into shadow zones by utilizing convergence zones. This simulation demonstrates that optimal aRIS placement at the sound channel axis depth eliminates shadow zones, enhancing deep-sea communication coverage.

Then we perform a quantitative evaluation of coverage capability, defining coverage capability as the proportion of spatial points within a region where the energy exceeds a certain threshold. This can be formulated as follows:
\begin{equation}
\eta_{coverage} = \frac{S_{\text{covered}}}{S_{\text{total}}}.
\end{equation}
where \(S_{\text{covered}}\) is the area of points with energy above the threshold, and \(S_{\text{total}}\) is the total area of the region. The energy threshold is determined using the transmission loss (TL) from the sonar equation\cite{komari2018passive}. By adjusting the source level (SL), we can reduce the TL threshold. We select common TL thresholds ranging from 100 to 150 dB and observed the coverage capability within the first shadow zone (20-30 km horizontal range, from the source depth to the seabed convergence zone). 

\begin{figure}[htbp]
\centerline{\includegraphics[width=1\textwidth]{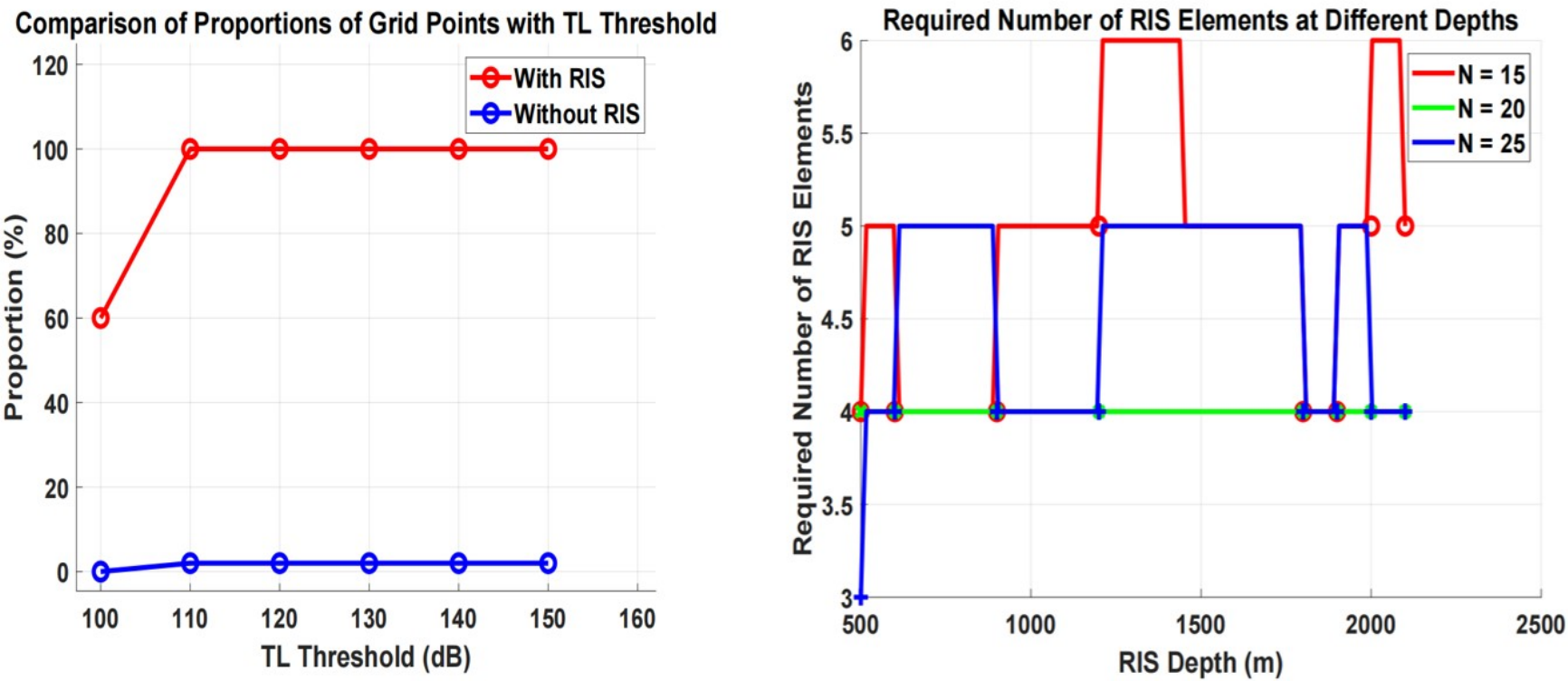}}
\caption{Comparison of Proportions of Grid Points with TL Threshold (Without and With RIS) (left); Required Number of RIS Units at Different Depths for Different Array Sizes (N = 15, 20, 25) (right).}
\label{fig:coverage_comparison}
\end{figure}

Note that increasing the SL corresponds to tolerating a higher TL, which means that the TL threshold will gradually increase. As illustrated on the left side of Fig. \ref{fig:coverage_comparison}, without the aRIS-assisted coverage, a significant portion of the region remains uncovered, even with the increased SL. The coverage proportion reaches only about 10\%, indicating that the higher source energy alone cannot effectively cover the shadow zone due to the characteristics of the marine environment. However, with optimal aRIS deployment at the sound channel axis, increasing the source energy achieves nearly 100\% seamless coverage with the assistance of the aRIS. 

\vspace{-3mm}
In Section III, we validate that the beamforming gain of each aRIS is limited. Consequently, deploying aRIS at different depths requires increasing the number of units to ensure robust energy coverage. As the energy received from the sound source attenuates, aRIS must provide sufficient gain to maintain strong coverage. By solving for the transmission loss (TL), we numerically calculate the minimum number of aRIS units needed at various depths. Figure \ref{fig:coverage_comparison} (right) shows the required number of aRIS units for different element configurations (\(N \times N\) with \(N=15, 20, 25\)). The results indicate that the aRIS depth has minimal impact on the required number of aRIS elements, particularly for \(N=20\), where the performance remains stable across all depths. For smaller (\(N=15\)) and larger (\(N=25\)) aRIS sizes, slight fluctuations are observed at certain depths, but overall differences remain minor. It suggests that selecting a moderate aRIS size, such as \(N=20\), is sufficient to ensure stable performance across various deployment depths, with further adjustments offering limited performance gains.

Although deploying the aRIS at the sound channel axis provides optimal coverage performance, the gain required from the aRIS also needs to increase significantly. In contrast, at shallower depths, the required gain from the aRIS is lower, effectively simulating an increase in source level (SL) and thereby reducing the TL threshold. However, as shown on the left side of Figure \ref{fig:coverage_comparison}, deploying aRIS at non-optimal depths results in a reduced maximum tolerable TL threshold, gradually decreasing the actual covered area. Therefore, the supplementary gain and deployment position of the aRIS represent a balance between maximizing coverage performance and managing the required gain.


\subsection{ Shallow-Sea Coverage Optimization}\label{BB}
For shallow-sea coverage optimization, Section IV-B presents the numerical solution method. Here, for comparative purposes, we explore the coverage distances at different depths through specific numerical examples. The sound speed distribution, shown in the left of Fig. \ref{fig:shallow_sea_coverage}, follows the linear SSP given by the function\cite{bradley1971propagation} as
\begin{equation}
c(z) = 
\begin{cases} 
1500 + 0.1883 \cdot z & \text{if } z \leq 60, \\
1551.902 - 0.6817 \cdot z & \text{if } z > 60. 
\end{cases}
\label{eq:shal}
\end{equation}
The surface layer width is 60 m, and the total shallow-sea depth is 120 m. For the shallow-sea aRIS-assisted coverage area, since the added dimension is only in the \( r \)-dimension, we can focus on the increased coverage distance \( r \) to characterize the coverage capability of the aRIS, i.e. $\eta_{coverage} = \frac{r_{\text{covered}}}{r_{\text{total}}}$. Therefore, we traverse different depths to observe the aRIS-assisted coverage capability, as shown on the left side of Fig. \ref{fig:shalowsea_optimal_coverage}. From the obtained results, it is evident that placing the aRIS at the maximum depth, i.e. , the seabed, achieves the greatest coverage distance. It indicates that the deeper the aRIS is placed, the more it can utilize the sound speed gradient to create larger reflection angles, maximizing the reflection distance and thus covering more shadow zones. 

\begin{figure}[htbp]
\centerline{\includegraphics[width=1\textwidth]{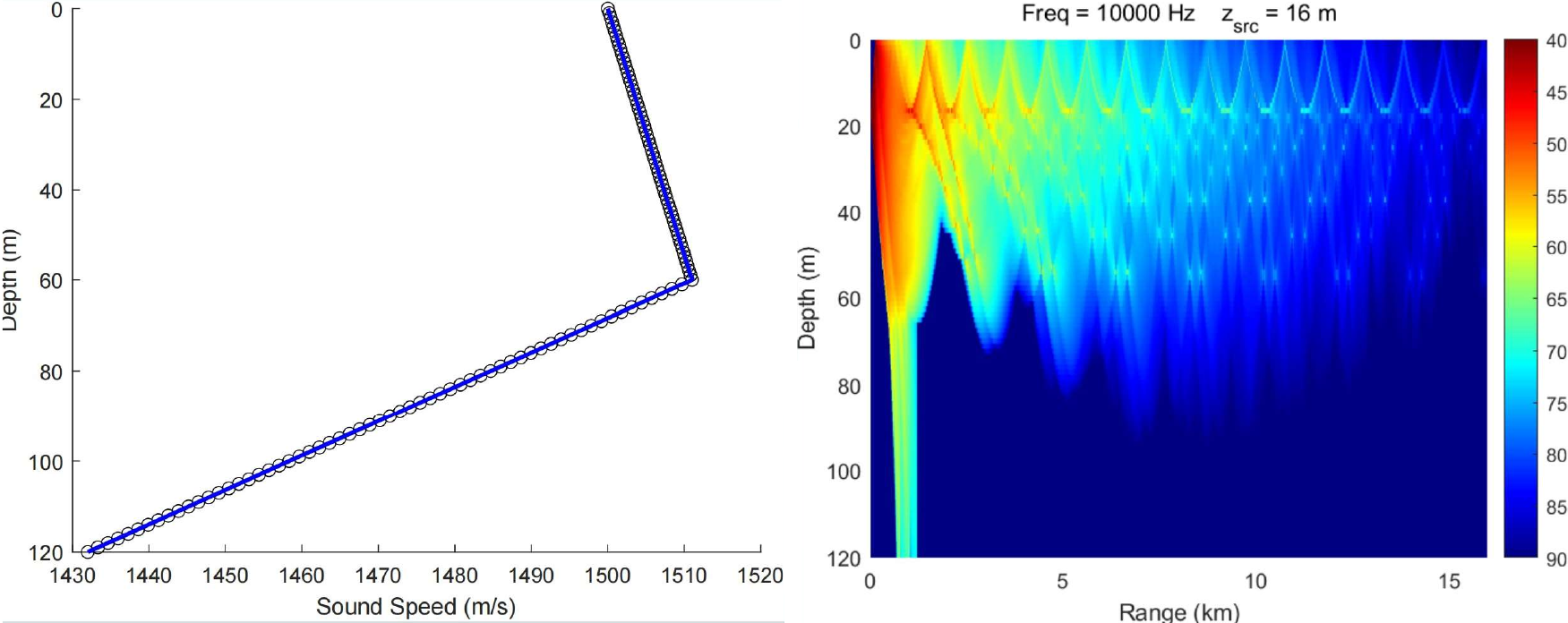}}
\caption{Standard shallow-sea sound speed profile model(left); Shallow-sea coverage model (right).}
\label{fig:shallow_sea_coverage}
\end{figure}


To achieve even greater coverage, a multi-hop relay approach using multiple underwater aRIS can be considered, as illustrated on the right side of Fig. \ref{fig:shalowsea_optimal_coverage}. This approach can further extend the coverage area and reduce the shadow zones. By numerically calculating the required number of aRIS at different depths to cover a specific area (10 km), we can determine the optimal number of aRIS units needed for full coverage. For optimal placement at the seabed depth, approximately 10 aRIS units are required, with each unit covering an average distance of 1 km. In contrast, for non-optimal placements, about 20 aRIS units are needed, doubling the number required for optimal placement. 
\vspace{-1mm} 
\begin{figure}[h]
\centerline{\includegraphics[width=1\textwidth]{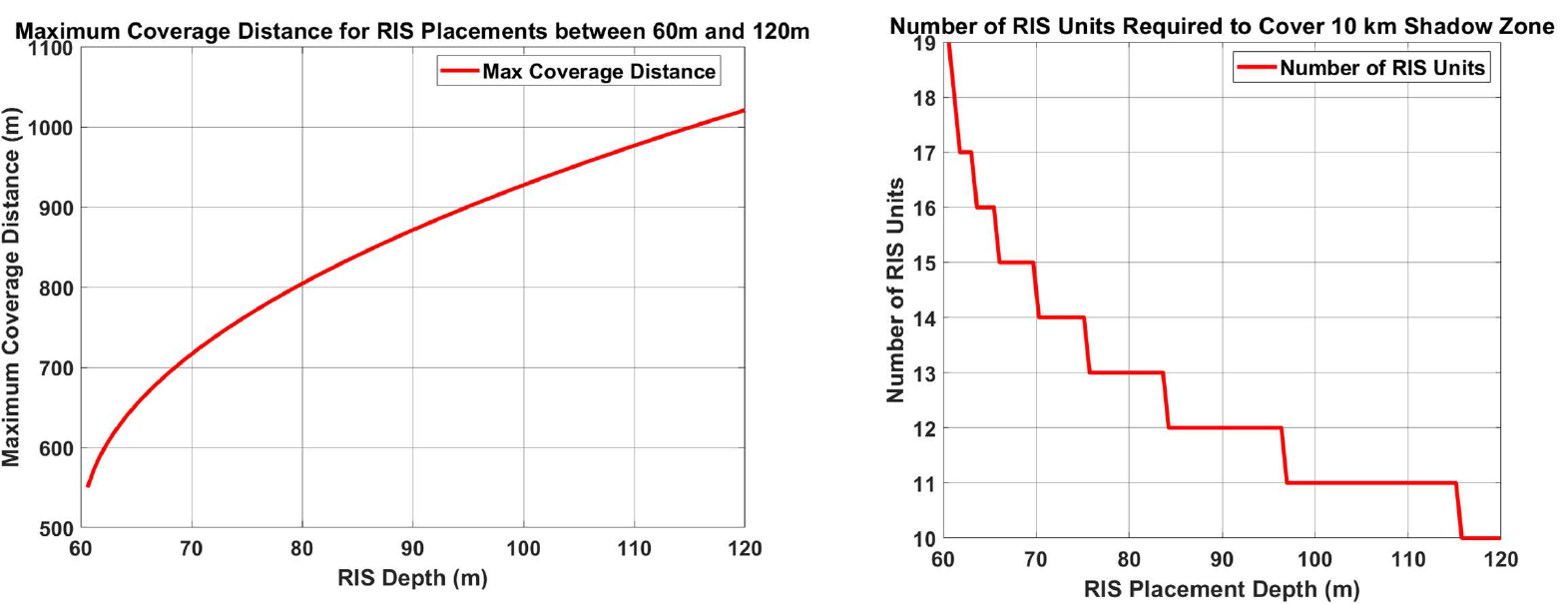}}
\vspace{-2mm} 
\caption{Coverage Proportion vs. RIS Placement Depth in Shallow Sea (left); Number of RIS Units Required to Cover a 10 km Shadow Zone at Different Depths in Shallow Sea (right).}
\label{fig:shalowsea_optimal_coverage}
\vspace{-2mm} 
\end{figure}

\vspace{-2mm} 
To compare coverage with and without aRIS, we simulated the energy distribution in a shallow-sea environment using Bellhop. The left side of Fig. \ref{fig:shallow_sea_coverage} shows that the numerically obtained optimal placement depth has practical significance. Deploying aRIS effectively reflects sound rays with large grazing angles that would otherwise be absorbed by the seabed, retaining energy within the shallow-sea sound channel and maximizing coverage. For even greater coverage, a multi-hop relay approach using multiple aRIS units can be considered, as illustrated on the right of Fig. \ref{fig:shallow_sea_coverage}, further extending the coverage area and reducing shadow zones.


\begin{figure}[htbp]
\centerline{\includegraphics[width=1\textwidth]{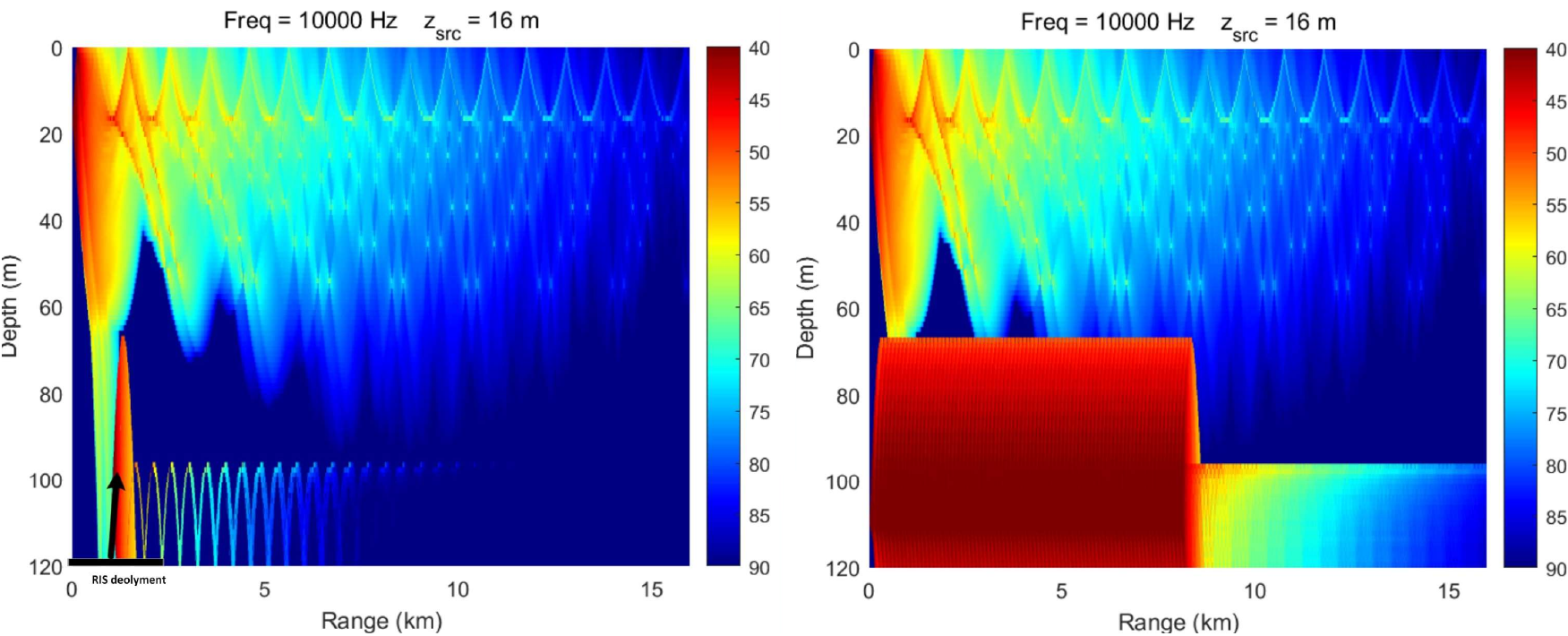}}
\caption{Energy Distribution with RIS Coverage(left); Multi-Hop Relay Approach with Multiple RIS (right).}
\label{fig:shallow_sea_coverage}
\end{figure}



\section{Impact Analysis of Dynamic Marine Environment and Corresponding Robust Solution}\label{Robust}

In this section, we analyze the stability and robustness of underwater aRIS deployment under dynamic ocean conditions. For shallow-sea scenarios, the aRIS platform can be secured to the seabed to prevent movement. However, in deep-sea scenarios, the aRIS is placed at the sound channel axis depth in mid-ocean\cite{du2019numerical}, where it may move vertically, horizontally, and rotate. Sec. V-A models these impacts, and Sec. V-B proposes robust solutions to ensure stable communication coverage despite environmental changes.

\subsection{Dynamic Marine Environment Impact Analysis}

\begin{figure}[htbp]
\centerline{\includegraphics[width=1\textwidth]{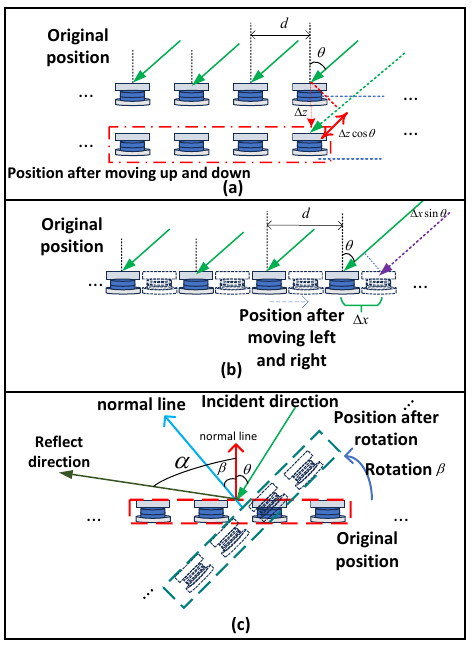}}
\caption{Impact of displacements and rotation on underwater acoustic RIS: (a) Vertical displacement, (b) Horizontal displacement, (c) Rotation.}
\label{fig:displacement_rotation}
\vspace{-3mm}
\end{figure}

Deploying underwater aRIS in real-world ocean environments subjects the platform to vertical and horizontal displacements, as well as array rotations. We analyze the impact of these factors on coverage performance, as illustrated in Fig.\ref{fig:displacement_rotation}. 

Vertical and horizontal displacements, caused by waves or currents, introduce signal arrival delays and affect phase synchronization, respectively. The phase deviations $\Delta \phi$ due to vertical displacement $\Delta z$ and horizontal displacement $\Delta x$ are given by:
\begin{equation}
\Delta \phi = k \Delta z \cos \theta, \quad \Delta \phi = k \Delta x \sin \theta,
\label{eq:displacement}
\end{equation}
where \(k\) is the wavenumber and \(\theta\) is the incident angle of the sound wave.

Platform rotation changes the normal of the aRIS array, affecting incident and reflection angles. If the array rotates by angle $\beta$, the phase compensation for the $n$-th array element needs to be adjusted to:
\begin{equation}
\Delta \phi = (n-1) k d \left( \sin(\theta + \beta) - \sin(\alpha - \beta) \right).
\label{eq:rotation}
\end{equation}

To maintain the original reflection direction in a dynamic environment, it is essential to accurately sense and estimate \(\Delta x\), \(\Delta z\), and the rotation angle \(\beta\).

\subsection{Corresponding Robust Solution}


For aRIS platform movements, accurate estimation and real-time correction are essential. Underwater acoustic systems like USBL can achieve centimeter-level precision \cite{rypkema2019passive}. Displacements \(\Delta x\) or \(\Delta z\) smaller than the array element spacing \(d\) require no compensation. For larger displacements, the terrestrial control center dynamically updates positional information to adjust phase compensation. Phase deviations are calculated as $
\Delta \phi_{\text{vertical}} = k \Delta z \cos \theta, \quad \Delta \phi_{\text{horizontal}} = k \Delta x \sin \theta.$



To address underwater aRIS rotation, gyroscopes estimate the rotation angle by measuring angular velocity \(\omega(t)\) and integrating over time to obtain \(\theta(t)\). The measured angular velocity includes errors:
\begin{equation}
\omega_{\text{measured}}(t) = (1 + \alpha) \omega(t) + b + n(t),
\end{equation}
where \(\alpha\) is the scale factor error, \(b\) is the bias error, and \(n(t)\) is random noise. The measured rotation angle is:
\begin{equation}
\beta_{\text{measured}}(t) = \int_{0}^{t} [(1 + \alpha) \omega(\tau) + b + n(\tau)] \, d\tau.
\end{equation}

For small angles:
\begin{equation}
\sin(\theta + \beta_{\text{measured}}) \approx \sin(\theta + \beta) + \Delta \beta \cos(\theta + \beta),
\end{equation}
where \(\Delta \beta = \alpha \beta + bt + \int_{0}^{t} n(\tau) \, d\tau\). Since \(\Delta \beta\) is small, its impact on phase compensation is negligible.

\subsubsection{Displacement Analysis}
\begin{figure}[htbp]
\centerline{\includegraphics[width=1\textwidth]{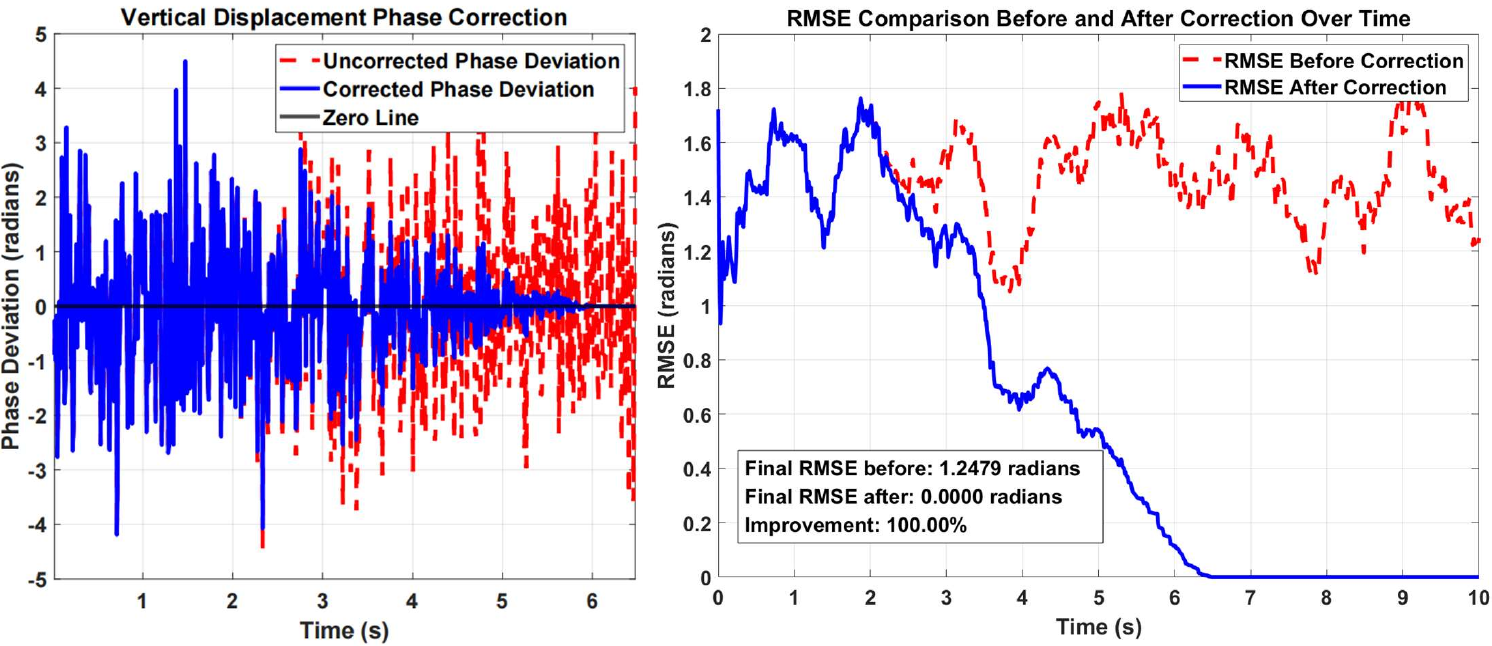}}
\vspace{-2mm}
\caption{Vertical Displacement Phase Correction over Time (left); Vertical RMSE Comparison Before and After Correction Over Time (right).}
\label{fig:vphase_deviation}
\end{figure}

\begin{figure}[htbp]
\centerline{\includegraphics[width=1\textwidth]{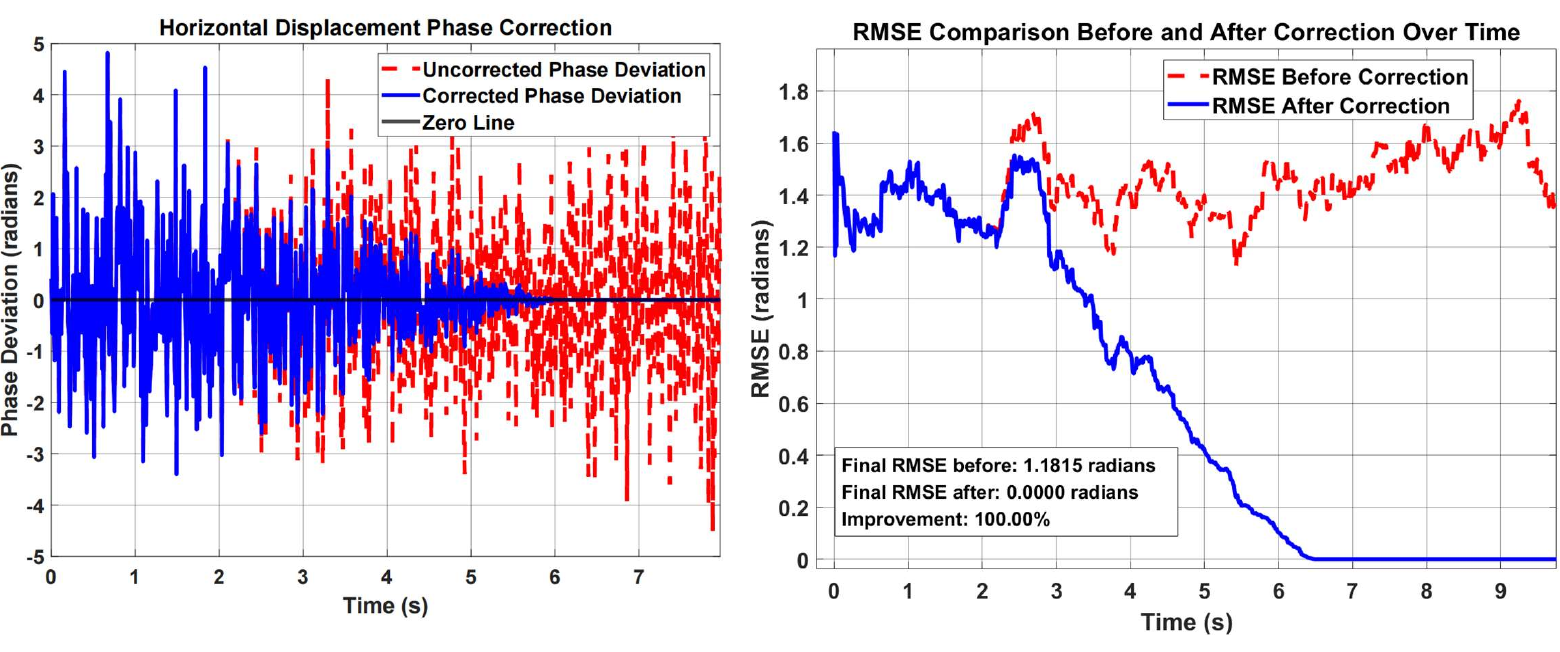}}
\vspace{-2mm}
\caption{Horizontal Displacement Phase Correction over Time (left); Horizontal RMSE Comparison Before and After Correction Over Time (right).}
\label{fig:hphase_deviation}
\end{figure}

Assume vertical and horizontal displacements of 50 cm with a standard deviation of 0.05 and an incident angle of 45 degrees. Key parameters include maximum displacement (\(\Delta z\) or \(\Delta x\)) of 0.5 m, standard deviation of 0.05, \(\theta = 45^\circ\), and element spacing of 0.075 m. Both displacements cause significant phase deviations, which are reduced to nearly zero after phase correction. Fig.~\ref{fig:vphase_deviation} and Fig.~\ref{fig:hphase_deviation} show the time-series phase deviation (red dashed for uncorrected, blue solid for corrected) and RMSE reduction, demonstrating a 99\% improvement after correction.

\begin{figure}[h]
\centerline{\includegraphics[width=1\textwidth]{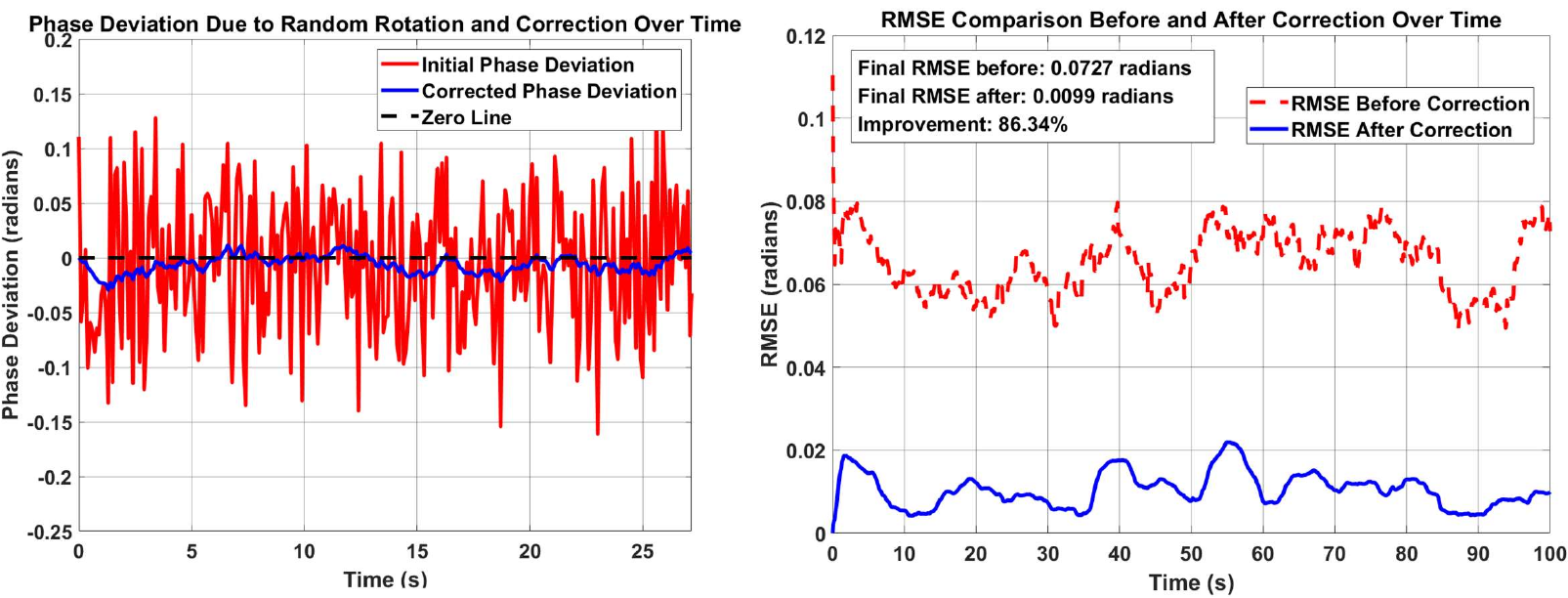}}
\vspace{-2mm}
\caption{Phase deviation due to random rotation and correction over time (left); Rotation RMSE Comparison Before and After Correction Over Time (right).}
\label{fig:rotation_phase_correction}
\end{figure}

\subsubsection{Rotation Analysis}

Using typical values\cite{ahmad2013reviews}, we assume a scale factor error \(\alpha \approx 0.001\), a bias error \(b \approx 0.01\) degrees/second, and random noise \(n(t) \approx 0.005\) degrees/second. For \(\beta \approx 5\) degrees and time \(t = 10\) seconds, we calculate \(\alpha\beta  = 0.001 \times 5 = 0.005\) degrees, \(bt = 0.01 \times 10 = 0.1\) degrees, and \(\int_{0}^{t} n(\tau) \, d\tau \approx 0.016\) degrees (assuming the standard deviation of white noise). The total error is:
$\Delta\beta \approx 0.005 + 0.1 + 0.016 = 0.121 \text{ degrees}.$
This error is within the acceptable range, indicating stable and effective rotational measurement compensation. The small angle approximation ensures that the error impact on phase compensation is minimal. As illustrated in Fig. \ref{fig:rotation_phase_correction}, the simulations demonstrate the phase correction process for rotation-induced errors. The corrected phase deviations approach zero over time, demonstrating the effectiveness of our compensation method. The RMSE improves by approximately 80\% after correction, highlighting the effectiveness of our phase correction method.

\section{Conclusions}\label{Con}
In this paper, we propose a novel approach for ubiquitous underwater communication network coverage using underwater aRIS to address the challenge of shadow zones. We conduct detailed modeling and analysis of shadow zones, establishing optimal coverage areas for aRIS deployment. Enhanced with practical engineering considerations, our aRIS prototype demonstrates directional reflection and enhanced signal coverage through beamforming in pool tests. For shallow-sea environments, deploying aRIS at the seabed redirects high-grazing-angle rays to cover shadow zones below the surface layer, while in deep-sea environments, positioning aRIS at the sound channel axis depth proves most effective. We analyze the impact of dynamic marine conditions on aRIS performance, proposing solutions to ensure robust deployment and effective reflection despite environmental changes. Comprehensive simulations validated through Bellhop-based models demonstrate that our underwater aRIS system significantly enhances communication reliability and coverage, indicating its potential for addressing underwater acoustic shadow zone communication challenges.

\section*{Acknowledgement}

The work of Longfei Zhao and Zhi Sun are supported by the National Natural Science Foundation of China, with a Grant of No. 62271284 for the project "Towards Acoustic Reconfigurable Intelligent Surface for High-data-rate Long-range Underwater Communications".

The work of Jingbo Tan and Jintao Wang are supported by the National Natural Science Foundation of China, with a Grant of No. 62401315 for the project "Acoustic Reconfigurable Intelligent Surface Based High-speed Mid-Long-Range Deep-Sea Underwater Acoustic Communication Theory and Key Technologies".

The work of I. F. Akyildiz is supported by the Icelandic Research Fund administered by Rannís – the Icelandic Centre for Research, with a Grant of Excellence No. 239994-051 for the project “HAF: Under-water Robotics Sensor Networks with Multi-Mode Devices and Remote Power Charging Capabilities.

\bibliographystyle{IEEEtran}
\bibliography{main}

\end{document}